\documentclass[eqsecnum,floats,aps,nofootinbib]{revtex4}
\usepackage{amsmath,amssymb,latexsym,epsfig}

\def\d{\mathrm{d}}
\def\e{\mathrm{e}}
\def\lbar{\mbox{\rlap{$\raise3pt\hbox{--}$}{$\lambda$}}}

\begin{document}
\thispagestyle{empty}

\title{Relativistic equations with fractional and pseudodifferential operators}

\author{D. Babusci}
\email{danilo.babusci@lnf.infn.it}
\affiliation{INFN - Laboratori Nazionali di Frascati, via E. Fermi, 40, 00044 
Frascati (Roma), Italy} 
\author{G. Dattoli}
\email{giuseppe.dattoli@enea.it}
\author{M. Quattromini}
\email{quattromini@frascati.enea.it}
\affiliation{ENEA -  Centro Ricerche Frascati, via E. Fermi, 45, 00044,  
Frascati (Roma), Italy} 

\begin{abstract}
In this paper we use different techniques from the fractional and pseudo-operators calculus to solve partial differential 
equations involving operators with non integer exponents. We apply the method to equations resembling 
generalizations of the heat equations and discuss the possibility of extending the procedure to the relativistic 
Schr\"odinger and Dirac equations.
\end{abstract}

\maketitle

\section{Introduction} \label{sec:intro}
Evolution equations involving fractional derivatives play an important role in the theory of transport in heterogeneous 
media \cite{Zasla} and different techniques have been developed \cite{Compte}, or are under study \cite{Bonil}, to deal, 
in an unambiguous way, with the underlying formalism.

This paper is a contribution to the theory of operators involving noninteger exponents and we will show how many of 
the techniques employed within such a context can be used to get further insight into the formalism of differential 
equations, like the relativistic Schr\"odinger equation or the pseudoheat equation. Here we will be mainly concerned with 
the mathematical aspects and postpone the application of the underlying formalism to a forthcoming investigation.

Different mathematical tools to deal with ÒfractionalÓ differential equations are available. Many of them were not 
considered by physicists during the second decade of the last century, at the time of the proposal of the Dirac equation. 
Presently some progress has been made toward the understanding of fractional operators and their use is not limited to 
pure mathematics anymore. However, it seems that a good convergence can be realized by at least three different points 
of view, namely the formalism of pseudodifferential operators, the fractional calculus and the theory of integral operators, 
as will be shown in this paper.

We will discuss evolution equations containing the square roots of differential operators and we will adopt different 
methods of solution, involving the techniques from fractional calculus, from the theory of pseudo-operators \cite{Taylo},  
and from that of the generalized integral transform \cite{DitPru}. In the following we will see that the use of a \emph{judicious} 
combination of operational techniques and of integral transform methods is one of the main ingredients to get a solution 
of partial differential equation involving fractional derivatives.

Evolution equations expressed in terms of the square root of an operator have a quite old story in physics and an example 
is provided by the debate developed around the meaning of the relativistic Schr\"odinger equation, which paved the way 
to the formulation of the Dirac equation for the electron \cite{Dira1}. As a first example of this class of equations, we consider, 
without mentioning the specific physical problems it describes, the following equation (the $x$ and $\tau$ variables are 
dimensionless):
\begin{equation}
\label{eq:sqrder}
\partial_\tau F(x,\tau) \,=\, -\,\partial_x^{1/2}\,F(x,\tau)\,, \qquad\qquad F(x,0) \,=\, f(x)\,.
\end{equation}
The relevant solution can formally be written as 
\begin{equation}
F(x,\tau) \,=\, \exp\left(-\tau\,\partial_x^{1/2}\right)\,f(x)\;,
\end{equation}
that, unless we specify the action of the exponential operator containing the square root of a derivative on the initial function $f(x)$, 
is just a way of restating Eq. \eqref{eq:sqrder}. The Doetsch transform \cite{Doets} (see Appendix A) is the appropriate tool to get 
a solution for our problem, that can be formally written as
\begin{eqnarray}
\label{eq:sqrsol}
F(x,\tau) \!\!&=&\!\! \frac{1}{2\,\sqrt{\pi}}\,\int_0^\infty\,\d t\,\frac{1}{t^{3/2}}\,\exp\left(-\,\frac{1}{4\,t} \,-\, 
                               t\,\tau^2\,\partial_x\right)\,f(x) \nonumber \\
                  &=&\!\!  \frac{1}{2\,\sqrt{\pi}}\,\int_0^\infty\,\d t\,\frac{1}{t^{3/2}}\,\exp\left(-\,\frac{1}{4\,t}\right)\,
                               f(x \,-\, \tau^2\,t)\,,
\end{eqnarray}
and is valid if the integral on the right-hand side converges. As we will see in the following, the fact that the solution of Eq. \eqref{eq:sqrder} 
is given by the form of an integral representation is a common feature of problems involving fractional derivatives \cite{OldSpa}.

We consider now the equation
\begin{equation}
\label{eq:psehea}
\partial_\tau F(x,\tau) \,=\, -\,\sqrt{1 \,-\,\partial_x^2}\,F(x,\tau)\,, \qquad\qquad F(x,0) \,=\, f(x)\,,
\end{equation}
to which we will refer as the pseudofractional (or simply pseudo) heat equation. By applying the evolution operator formalism, 
the solution of this equation can be written as follows
\begin{equation}
F(x,\tau) \,=\, \e^{-\tau\,\sqrt{1 \,-\,\partial_x^2}}\,f(x)\;,
\end{equation}
and, according to the Doetsch transform, we get
\begin{eqnarray}
\label{eq:pseFxt}
F(x,\tau) \!\!&=&\!\! \frac{1}{2\,\sqrt{\pi}}\,\int_0^\infty\,\d t\,\frac{1}{t^{3/2}}\,\exp\left[-\,\frac{1}{4\,t} \,-\, 
                               t\,\tau^2\,(1 \,-\,\partial_x^2)\right]\,f(x) \nonumber \\
                  &=&\!\!  \frac{1}{2\,\sqrt{\pi}}\,\int_0^\infty\,\d t\,\frac{1}{t^{3/2}}\,
                               \exp\left(-\,\frac{1}{4\,t} \,-\, t\,\tau^2\right)\,e^{t\,\tau^2\,\partial_x^2}\,f(x)\,.
\end{eqnarray}

The crucial step to achieve the complete solution consists in specifying the action of the exponential involving the second-order 
derivative. This can be done by using the Gauss transform \cite{Datto}
\begin{equation}
\e^{\alpha\,\partial_x^2}\,f(x) \,=\, \frac{1}{2\,\sqrt{\pi\,\alpha}}\,\int_{-\infty}^\infty\,\d \xi\,
                                                    \exp\left\{-\,\frac{(x\,-\,\xi)^2}{4\,\alpha}\right\}\,f(\xi)\,,
\end{equation}
that allows us to write
\begin{equation}
\label{eq:psesol}
F(x,\tau) \,=\, \frac{1}{4\,\sqrt{\pi}\,\tau}\,\int_0^\infty\,\d t\,\frac{1}{t^2}\,\exp\left(-\,\frac{1}{4\,t} \,-\, 
                               t\,\tau^2\right)\,\int_{-\infty}^\infty\,\d \xi\,\exp\left\{-\,\frac{(x\,-\,\xi)^2}{4\,t\,\tau^2}\right\}\,f(\xi)\,.
\end{equation}
As a particular case, we note that if $f(x) = e^{-x^2}$ we can use the so called Glaisher identity \cite{Datt1}
\begin{equation}
\label{eq:glais}
e^{\alpha\,\partial_x^2}\,e^{-x^2} \,=\, \frac{1}{\sqrt{1\,+\,4\,\alpha}}\,\exp\left(-\,\frac{x^2}{1\,+\,4\,\alpha}\right)\,,
\end{equation}
that, once inserted in Eq. \eqref{eq:pseFxt}, provides the result
\begin{equation}
F(x,\tau) \,=\, \frac{1}{2\,\sqrt{\pi}}\,\int_0^\infty\,\d t\,\frac{1}{t^{3/2}\,\sqrt{1\,+\,4\,t\,\tau^2}}\,
                      \exp\left\{-\,\left(\frac{1}{4\,t} \,+\, t\,\tau^2 \,+\, \frac{x^2}{1\,+\,4\,t\,\tau^2}\right)\right\}\,.
\end{equation}

\begin{figure}[htb]
\centering
\includegraphics[height=7cm]{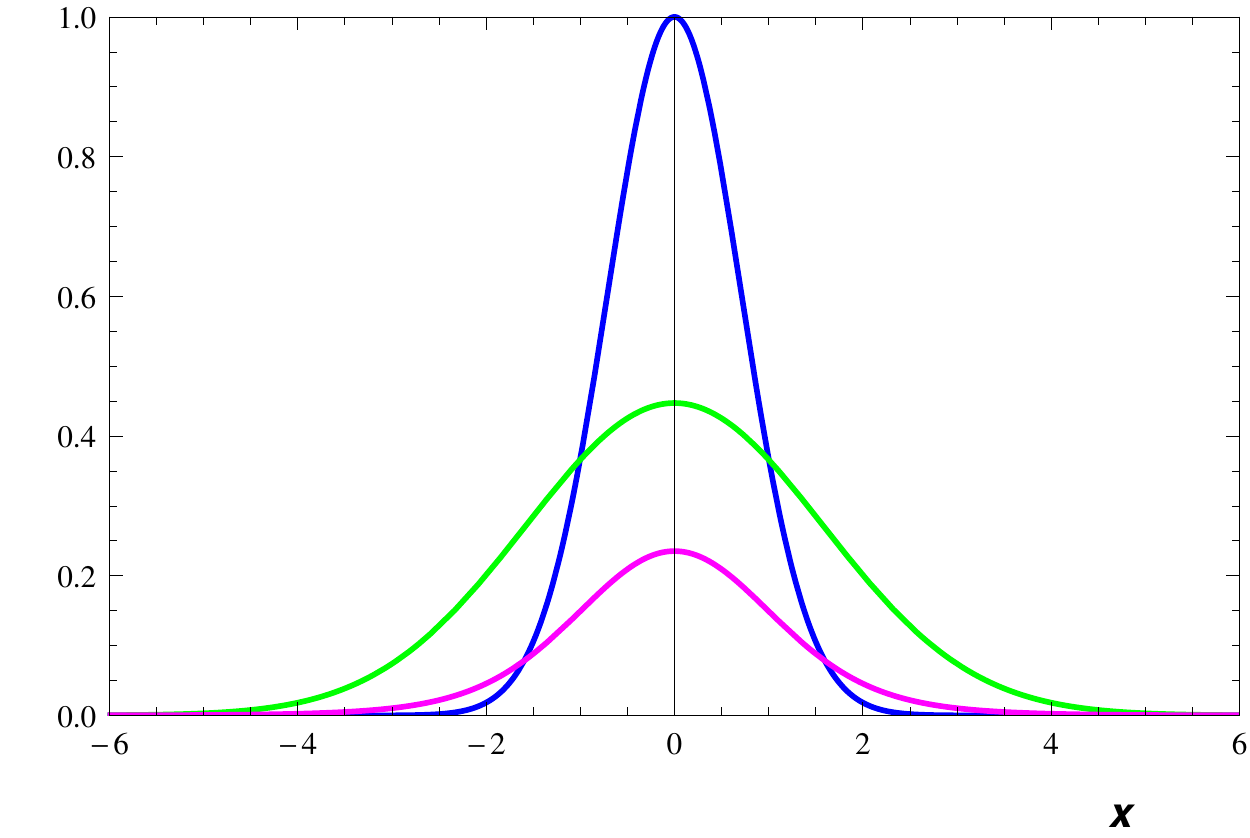}
\caption{The evolution of a packet initially Gaussian (blue) according to an ordinary (green) 
              and a fractional (purple) heat equation. The case considered corresponds to $\tau = 1$.}
\label{fig:cmp1}
\end{figure}

As clearly shown in Fig. \ref{fig:cmp1}, compared to the ordinary heat diffusion equation, the pseudoheat equation yields an 
evolution characterized by a less significant spreading and a larger reduction of the peak. 

More conventional methods, based, for example, on Fourier transform techniques can also be used. If we assume that the function 
$f(x)$  admits a Fourier transform, namely
\begin{equation}
\label{ }
f(x) \,=\, \frac{1}{\sqrt{2\,\pi}}\,\int_{-\infty}^\infty\,\d k\,\tilde{f}(k)\,e^{i\,k\,x}
\end{equation}
the solution of our problem can be obtained as\footnote{In the second line we have used the identity
$f(\partial_x)\,\e^{\alpha\,x} \,=\, f(\alpha)\,\e^{\alpha\,x}$.}
\begin{eqnarray}
\label{eq:psefou}
F(x,\tau) \!\!&=&\!\! \frac{1}{\sqrt{2\,\pi}}\,\e^{-\,\tau\,\sqrt{1\,-\,\partial_x^2}}\,
                               \int_{-\infty}^\infty\,\d k\,\tilde{f}(k)\,e^{i\,k\,x} \nonumber \\
                   &=&\!\! \frac{1}{\sqrt{2\,\pi}}\, \int_{-\infty}^\infty\,\d k\,
                                \e^{-\,\tau\,\sqrt{1\,+\,k^2}}\,\tilde{f}(k)\,e^{i\,k\,x}\,.
\end{eqnarray}
Solutions \eqref{eq:psesol} and \eqref{eq:psefou} are both expressed in terms of an integral transform and, in principle, there is no 
reason to prefer one over the other except for the nature of the initial function and the convergence of the integral involved in the 
expressions. 

The use of the Fourier transform method is a useful alternative to the method we are discussing when the pseudo-operator is not a 
function of the derivative operator only. To clarify this point we note that the evolution equation
\begin{equation}
\label{eq:foutran}
\partial_\tau\,F(x,\tau) \,=\, P(\partial_x)\,F(x,\tau)\;, \qquad\qquad F(x,0) \,=\, f(x)\,,
\end{equation}
with $P(.)$ an operator or an analytical function, can be solved as
\begin{equation}
F(x,\tau) \,=\, \frac{1}{\sqrt{2\,\pi}}\,\int_{-\infty}^\infty\,\d k\,e^{\tau\,P(i\,k)}\,\tilde{f}(k)\,e^{i\,k\,x}\,, 
\end{equation}
which holds if the integral converges. On the contrary, if  $P(.)$ is a function of the derivative operator and the coordinate, this method, 
even though still applicable (at the price of using a more complicated family of ``symbols''),  is more cumbersome and an alternative 
procedure might be more convenient. This is the case of the following equation
\begin{equation}
\partial_\tau\,F(x,\tau) \,=\, -\,\sqrt{x \,-\,c\,\partial_x}\,F(x,\tau)\;, \qquad\qquad F(x,0) \,=\, f(x)\,,
\end{equation}
where $c$ is a constant. Here, the use of Doetsch transform and Weyl disentanglement formula\footnote{If [$\hat{A},\hat{B}] = k \in \mathbb{C}$, 
then $\e^{\hat{A}\,+\,\hat{B}}\,=\,\e^{-\,k/2}\,\e^{\hat{A}}\,\e^{\hat{B}}$ (see Ref. \cite{Datto} for details).}  allow us to write the relevant solution 
in a quite straightforward way as follows
\begin{eqnarray}
F(x,\tau) \!\!&=&\!\! \frac{1}{2\,\sqrt{\pi}}\,\int_0^\infty\,\d t\,\frac{1}{t^{3/2}}\,
                               \exp\left\{-\,\frac{1}{4\,t} \,-\, t\,\tau^2\right\}\,\e^{-\,t\,\tau^2\,(x \,-\, c\,\partial_x)}\,f(x) \\
                   &=&\!\! \frac{1}{2\,\sqrt{\pi}}\,\int_0^\infty\,\d t\,\frac{1}{t^{3/2}}\,
                                \exp\left\{-\,\left(\frac{1}{4\,t} \,+\, t\,\tau^2 \,+\, \frac{\tau^4\,t^2}{2}\right)\right\}\,
                                \e^{-\,t\,\tau^2\,x}\,f(x \,+\, c\,\tau^2\,t)\,, \nonumber
\end{eqnarray}
while the use of pseudodifferential operators would require a more cumbersome analysis.

According to the discussion developed so far it is evident that the combined use of integral transform and operational methods offers 
the natural environment to deal with fractional and/or pseudo-operators. Within this context  the Laplace  transform method offers a wealth 
of possibilities. For example, by assuming that the identity \cite{Doets}
\begin{equation}
\frac{1}{\sqrt{a^2 \,+\, b^2}} \,=\, \int_0^\infty\,\d t\,J_0(b t)\, \e^{-\,a\,t}
\end{equation}
can be extended to operators, we find \cite{Datt2}
\begin{equation}
\frac{1}{\sqrt{\partial_x^2 \,+\, 1}} \,=\, \int_0^\infty\,\d t\,J_0(t)\, \e^{-\,t\,\partial_x}\;.
\end{equation}
We can therefore express the solution of a problem of the type
\begin{equation}
\sqrt{\partial_x^2 \,+\, 1}\,f(x) \,=\, g(x)\,,
\end{equation}
with $f(x)$ an unknown function, as
\begin{equation}
f(x) \,=\, \int_0^\infty\,\d t\,J_0(t)\,g(x \,-\,t)\;.
\end{equation}
The use of the above transform allows to cast the equation
\begin{equation}
\partial_\tau\,F(x,\tau) \,=\, \sqrt{\partial_x^2 \,+\, 1}\,F(x,\tau)\;,
\end{equation}
in the form
\begin{equation}
\label{eq:transF}
F(x,\tau) \,=\, \int_0^\infty\,\d t\,J_0(t)\,\partial_\tau\,F(x \,-\,t, \tau)\;.
\end{equation}                  
Unfortunately, this equation is of limited usefulness since the convolution integral on right-hand side does not produce well behaved functions.

\section{Relativistic Schr\"odinger equation} \label{sec:relscho}
As a variation on the pseudo-fractional heat equation, we consider the relativistic (1+1 dimensional) Schr\"odinger equation for a particle 
of mass $m$, that writes
\begin{equation}
\label{eq:schro}
i\,\partial_\tau\,\Psi(\eta,\tau) \,=\, \sqrt{1 \,-\, \partial_\eta^2}\,\Psi(\eta,\tau)\;, \qquad\qquad \Psi(\eta,0) \,=\, \phi(\eta),
\end{equation}
where time and position variables are normalized to the Compton wavelength ($\lbar_c = \hbar/mc$): 
\begin{equation}
\tau \,=\, \frac{c\,t}{\lbar_c}\,, \qquad \eta \,=\, \frac{x}{\lbar_c}\,. \nonumber
\end{equation}

It describes the quantum evolution of a relativistic free particle and does not include negative energy contributions. The use of the Schr\"odinger 
equation in the form given by Eq. \eqref{eq:schro} has been criticized on the basis of arguments concerning its nonlocality, as a 
consequence of the nonlocal character of the fractional operators. The problem is of genuine mathematical nature and does not arise from the 
potential as in the case of the nonlocalities associated with the fermionic nature of nucleons in nuclear many-body problems. The initial negative 
reaction to Eq. \eqref{eq:schro} stems, probably, from the intrinsic mathematical difficulties rather than due to a real physical reason.  
In Ref. \cite{BjoDr} it is clearly stated that Eq. \eqref{eq:schro} has been ruled out as a tool to develop the relativistic quantum theory for reason of 
simplicity (i.e., to deal with simpler expressions, avoiding square roots of operators) but not on the basis of a real ``physical cogency.''
Even though the questions about the intrinsic nonlocal nature of Eq. \eqref{eq:schro} are subtle and have opened a long-standing debate, the 
relevant physical consequences do not display any disturbing feature. This point has been reconsidered in various papers (see 
Ref. \cite{Lamme}) where its physical legitimacy has been discussed in depth.

From the mathematical point Eq. \eqref{eq:schro} can be transformed into a pseudoheat equation by a Wick rotation, and, thus, its solution can be 
written by following the procedure outlined in Sec. \ref{sec:intro}. However, the obtained result is ambiguous and hampered by the relevant physical 
interpretation and the fact that the convergence of the integral representation \eqref{eq:psesol} with $\tau^2 \to -\tau^2$  is not ensured. For 
this reason we will treat the problem following a different approach. 

\begin{figure}[htb]
\centering
\includegraphics[height=7cm]{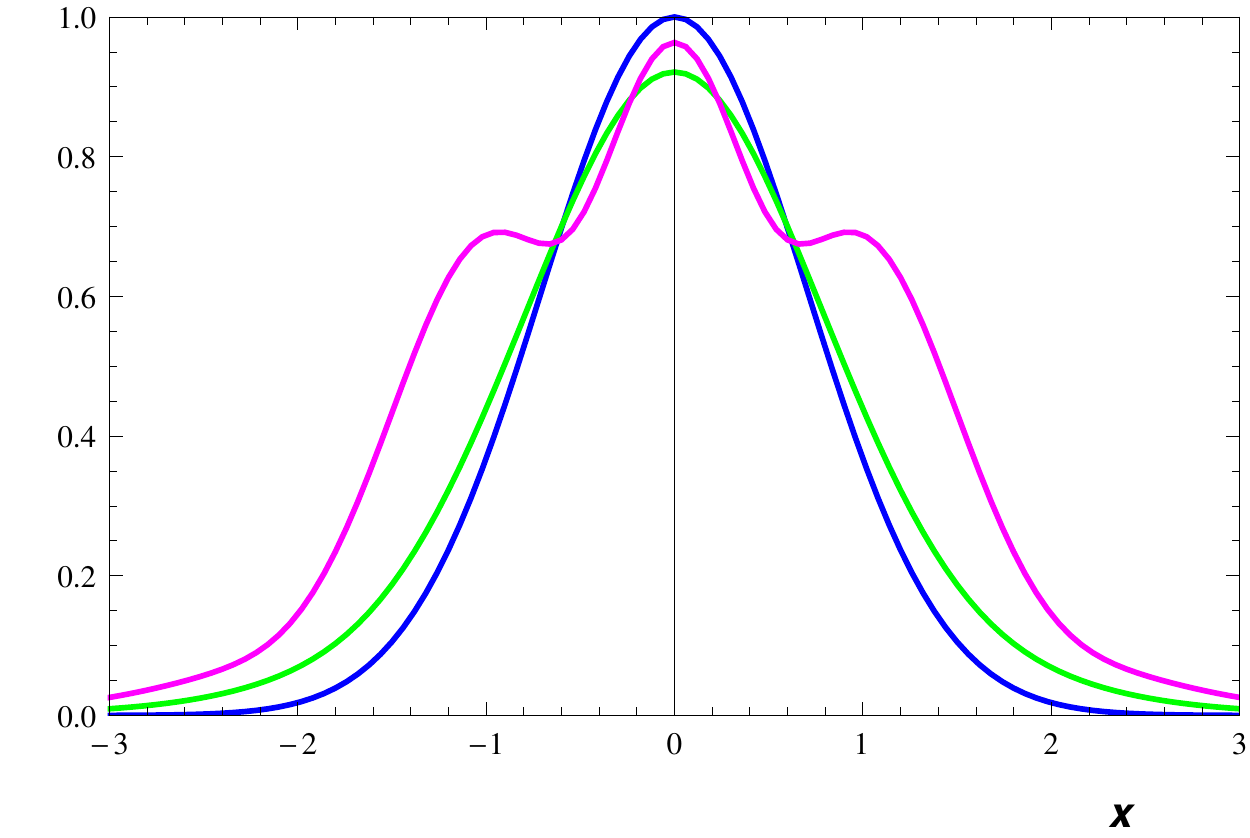}
\caption{Time evolution of an initially Gaussian packet (blue) undergoing  
              diffusion ruled by a relativistic Schr\"odinger equation: $\tau = 0.5$ (green) and 
              $\tau = 1$ (purple).}
\label{fig:relschro} 
\end{figure}

The  evolution operator associated to Eq. \eqref{eq:schro} can be written as follows
\begin{equation}
\hat{U} (\tau) \,=\, \e^{-\,i\,\tau\,\sqrt{1 \,-\, \partial_\eta^2}} \,=\,\sum_{n = 0}^\infty\,(-1)^n\,
                             \frac{\left[\tau^2\,(1 \,-\, \partial_\eta^2)\right]^n}{(2 n)!}\,\left\{1 \,-\, 
                             \frac{i\,\tau}{2n + 1}\,\sqrt{1 \,-\, \partial_\eta^2}\right\}\,.
\end{equation}
The fractional operator appearing in the series can be handled by means of a trick, often exploited in the theory of the fractional derivatives 
(see also the forthcoming section), namely 
\begin{equation}
\sqrt{1 \,-\, \partial_\eta^2}\,\phi(\eta) \,=\, \frac{1 \,-\, \partial_\eta^2}{\sqrt{1 \,-\, \partial_\eta^2}}\,\phi(\eta) \,=\,
\frac{1 \,-\, \partial_\eta^2}{\sqrt{\pi}}\,\int_0^\infty\,\d s\,\frac{\e^{-\,s}}{\sqrt{s}}\,\e^{s\,\partial_\eta^2}\,\phi(\eta),
\end{equation}
where the last equality has been obtained using the Laplace transform identity
\begin{equation}
\label{eq:traide}
a^{-\,\nu} \,=\,\frac{1}{\Gamma(\nu)}\,\int_0^\infty\,\d s\,\e^{-a\,s}\,s^{\nu - 1}\, \qquad\qquad (\nu > 0). 
\end{equation}
In the case $\phi(\eta) = \e^{-\,\eta^2}$, from the Glaisher identity \eqref{eq:glais} we obtain
\begin{equation}
\label{eq:phiet}
\sqrt{1 \,-\, \partial_\eta^2}\,\e^{-\,\eta^2} \,=\, \frac{1 \,-\, \partial_\eta^2}{\sqrt{\pi}}\,\int_0^\infty\,\d s\,
\frac{1}{\sqrt{s\,(1 \,+\, 4\,s)}}\,\exp\left\{-\,\left(s \,+\, \frac{\eta^2}{1 \,+\, 4\,s}\right)\right\}\;,
\end{equation} 
and taking into account the following identity involving the two-variable Hermite polynomials \cite{Datt2,Datt3}
\begin{equation}
\partial_x^n\,\e^{a\,x^2} \,=\, H_n(2\,a\,x,a)\,\e^{a\,x^2} \qquad\qquad 
\left(H_n (x,y) \,=\, n!\,\sum_{k = 0}^{[n/2]}\,\frac{x^{n \,-\,2\,k}\,y^k}{(n \,-\, 2\,k)!\,k!}\right)\,,
\end{equation}
we can cast the free particle solution of the relativistic Schr\"odinger equation in the form
\begin{equation}
\Psi (\eta,\tau) \,=\, A(\eta,\tau)\,\e^{-\,\eta^2} \,+\, i\,B(\eta,\tau),
\end{equation}
where
\begin{eqnarray}
 A(\eta,\tau) \!\!&=&\!\! \sum_{n = 0}^\infty\,(-1)^n\,\frac{\tau^{2 n}}{(2 n)!}\,\sum_{k = 0}^n\,
                                    (-1)^k\,\binom{n}{k}\,H_{2k} (2 \eta,-1)\,, \nonumber \\
 B(\eta,\tau) \!\!&=&\!\! \sum_{n = 0}^\infty\,(-1)^{n + 1}\,\frac{\tau^{2 n + 1}}{(2 n + 1)!}\,\sum_{k = 0}^{n + 1}\,
                                    (-1)^k\,\binom{n + 1}{k}\,f_{2k} (\eta)\,,  \\
f_{2 k} (\eta) \!\!&=&\!\! \frac{1}{\sqrt{\pi}}\,\int_0^\infty\,\d s\,\frac{\e^{-\,s}}{\sqrt{s\,(1 \,+\, 4\,s)}}\,
                                     H_{2 k} \left(\frac{2\,\eta}{1 \,+\,4\,s},-\,\frac{1}{1 \,+\,4\,s}\right)\,
                                     \exp\left\{-\,\frac{\eta^2}{1 \,+\, 4\,s}\right\}\;.\nonumber
\end{eqnarray}
The comparison of the previous solution with its non-relativistic counterpart shows significant differences, which becomes more evident with 
increasing time (see Fig. \ref{fig:relschro}). We will discuss the relevant physical meaning in the forthcoming sections.

Even though the above relation provides us with a well behaved solution, extendable to non-Gaussian packets (see the concluding section), 
its unappealing features stems from the fact that it is essentially a Taylor series expansion in time. To overcome this problem different 
strategies (including the already quoted Fourier transform method) will be developed in the forthcoming sections.

\begin{figure}[htb]
\centering
\includegraphics[height=7cm]{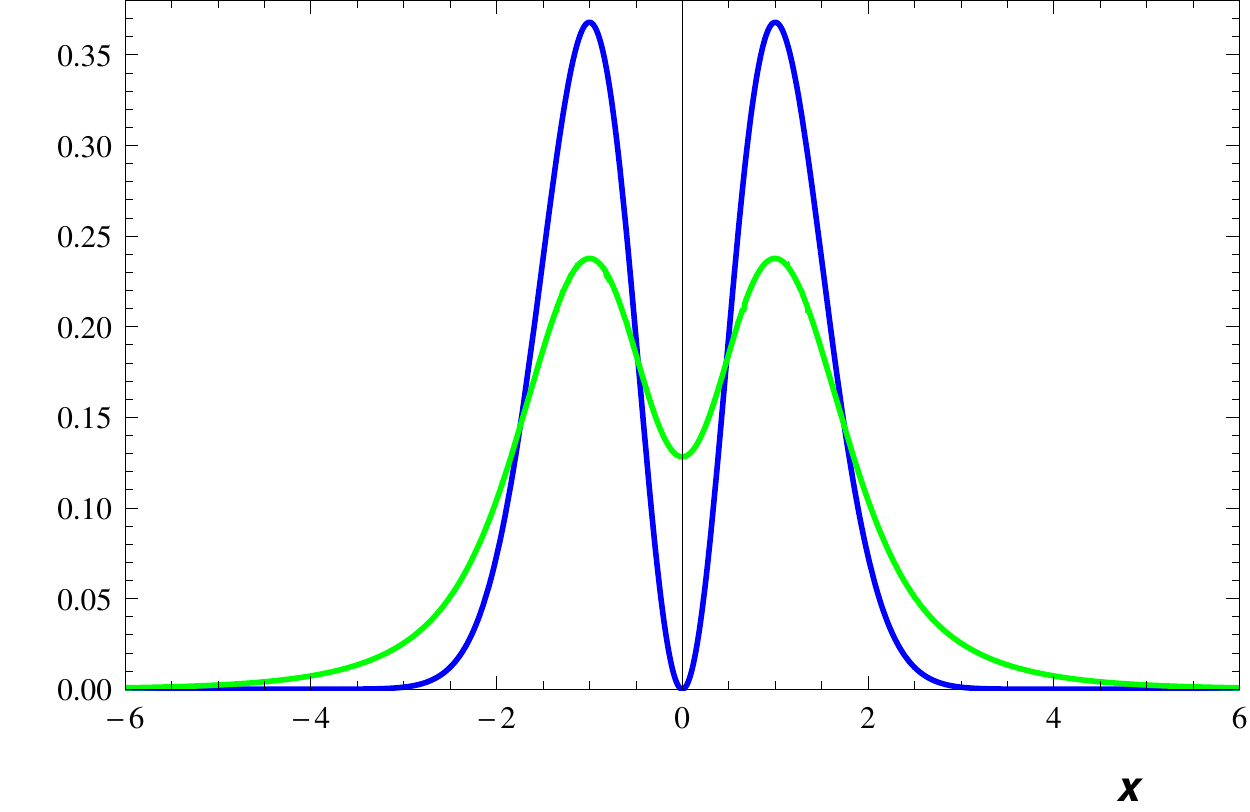}
\caption{The function $\Phi$ of Eq. \eqref{eq:solsbar} (green) calculated for an initial wave function 
$\bar{\Psi} (\xi, 0) = \xi^2\,\e^{-\,\xi^2}$ (blue).}
\label{fig:transf} 
\end{figure}

Another commonly adopted method in fractional calculus, which allows the elimination of noninteger exponents, is based on the elementary 
Laplace transform identity already exploited to derive Eq. \eqref{eq:phiet}. By using this identity, we can recast Eq. \eqref{eq:schro} in the form
\begin{equation}
\label{eq:schbar}
i\,\partial_\tau\,\bar{\Psi} (\eta,\tau) \,=\, -\,\partial_\eta^2\,\Phi (\eta,\tau)\,,
\end{equation}
where
\begin{eqnarray}
\label{eq:solsbar}
\bar{\Psi} (\eta,\tau) \!\!&=&\!\! \exp\left(-\,i\,\frac{\tau}{\sqrt{1 \,-\, \partial_\eta^2}}\right)\,\Psi (\eta,\tau) \nonumber \\
        \Phi (\eta,\tau) \!\!&=&\!\! \frac{1}{\sqrt{\pi}}\,\int_0^\infty\,\d s\,\frac{\e^{-\,s}}{\sqrt{s}}\,
                                                \e^{s\,\partial_\eta^2}\,\bar{\Psi} (\eta,\tau) \\ \nonumber
                                    &=&\!\! \frac{1}{2\,\pi}\,\int_0^\infty\,\d s\,\frac{\e^{-\,s}}{s}\,\int_{-\infty}^\infty\,\d \xi\,
                                                 \exp\left\{-\,\frac{(\eta \,-\, \xi)^2}{4\,s}\right\}\,\bar{\Psi} (\xi,\tau)\,. \nonumber          
\end{eqnarray}
This is a Schr\"odinger equation in which time and spatial derivative operators do not act on the same wave function. The action of the second-order 
spatial derivative is indeed mediated by another operator, inducing a kind of diffusion, whose effect is that of delocalizing the initial wave 
function, as  is illustrated in Fig. \ref{fig:transf}, where we have reported how the integral transform in the previous equation affects a given function. 
An alternative way of writing Eq. \eqref{eq:schbar} is the following
\begin{equation}
i\,\partial_\tau\,\bar{\Psi} (\eta,\tau) \,=\, -\,\partial_\eta^2\,\hat{D}\,\bar{\Psi} (\eta,\tau)\,,
\end{equation}
with
\begin{equation}
\hat{D} \,=\, \frac{1}{\sqrt{\pi}}\,\int_0^\infty\,\d s\frac{\e^{-\,s}}{\sqrt{s}}\,\e^{s\,\partial_\eta^2}\,,
\end{equation}
that allows the series solution 
\begin{equation}
\bar{\Psi} (\eta,\tau) \,=\, \sum_{n = 0}^\infty\,\frac{(i\,\tau)^n}{n!}\,\bar{\Psi}_n (\eta)
\end{equation}                           
where
\begin{equation}
\bar{\Psi}_n (\eta) \,=\, \partial_\eta^2\,\hat{D}\,\bar{\Psi}_{n - 1} (\eta) \,=\, \frac{1}{2\,\pi}\,
                                    \int_0^\infty\,\d s\,\frac{\e^{-\,s}}{s}\,\int_{-\infty}^\infty\,\d \xi\,
                                     \exp\left\{-\,\frac{(\eta \,-\, \xi)^2}{4\,s}\right\}\,\partial_\xi^2\,\bar{\Psi}_{n - 1} (\xi)\,.
\end{equation}

Let us now consider the inclusion of a potential, so that Eq. \eqref{eq:schro} becomes
\begin{equation}
i\,\partial_\tau\,\Psi (\eta,\tau) \,=\, \left[\sqrt{1\,-\,\partial_\eta^2} \,+\, V(\eta)\right]\,\Phi (\eta,\tau)\,.
\end{equation}
In this case, by introducing the transformation (a kind of interaction picture)
\begin{equation}
\Lambda (\eta,\tau) \,=\, \e^{-\,i\,\tau\,\sqrt{1\,-\,\partial_\eta^2}}\,\Psi (\eta,\tau)\,,
\end{equation}
the problem to solve becomes
\begin{equation}
\label{eq:schpot}
i\,\partial_\tau\,\Lambda (\eta,\tau) \,=\, \hat{V} (\eta,\tau)\,\Lambda (\eta,\tau)
\end{equation}
where
\begin{eqnarray}
\hat{V} (\eta,\tau) \!\!&=&\!\! \e^{i\,\tau\,\sqrt{1\,-\,\partial_\eta^2}}\,V (\eta)\, 
                                            \e^{-\,i\,\tau\,\sqrt{1\,-\,\partial_\eta^2}} \nonumber \\
                                &=&\!\! V(\eta) \,+\, i\,\tau\,\left[\sqrt{1\,-\,\partial_\eta^2}, V(\eta)\right] \,-\, \frac{\tau^2}{2}\,
                                            \left[\sqrt{1\,-\,\partial_\eta^2},\left[\sqrt{1\,-\,\partial_\eta^2}, V(\eta)\right]\right] \,+\, 
                                            O(\tau^3)\,. \nonumber
\end{eqnarray}                 
The previously outlined procedures can still be applied, but the solution of problem \eqref{eq:schpot} should be expressed as it follows:
\begin{equation}
\Psi (\eta,\tau) \,=\, \e^{i\,\tau\,\sqrt{1\,-\,\partial_\eta^2}}\,\left\{\,\exp\left(-\,i\,\int_0^\tau\,\d \tau^\prime\, 
                               \hat{V} (x,\tau^\prime)\right)\,\right\}_+ \Psi (\eta,0)\,,
\end{equation}
where $\{ \}_+$ denotes the Dyson time-ordering operation. In this case the ordering procedure becomes mandatory because the Hamiltonian 
contains the explicitly time-dependent term $\hat{V} (\eta,\tau)$, which does not commute with itself at different times.

\section{The Heisenberg picture and the fractional operators}\label{sec:heise}
In this section we will show how the use of methods involving fractional derivatives can be useful to study physical states whose evolution is  
ruled by relativistic Hamiltonians. We will show that these methods find a natural application in the treatment of the Heisenberg equations of 
motion of the physical observables and in the evaluation of their average values.

The relativistic Hamiltonian for a free particle is (we limit ourselves to the one-dimensional motion)
\begin{equation}
\label{eq:relham}
\hat{H} \,=\, c\,\sqrt{m^2\,c^2 \,+\, \hat{p}^2}\,,
\end{equation}
and the Heisenberg equations of motion for the position and momentum operators can be written as\footnote{Note that 
$[\hat{x}, f(\hat{p})] \,=\, i\,\hbar\,\partial_p\,f(\hat{p})$.}
\begin{equation}
\frac{\d}{\d t}\,\hat{x} \,=\, \frac{1}{i\,\hbar}\,\left[\hat{x},\hat{H}\right] \,=\, c\,\frac{\hat{p}}{\sqrt{m^2\,c^2 \,+\, \hat{p}^2}}\,
\qquad\qquad \frac{\d}{\d t}\,\hat{p} \,=\, 0\,,
\end{equation}
and the relevant solution reads
\begin{equation}
\label{eq:xoper}
\hat{x} (t) \,=\, \hat{x} (0) \,+\, \frac{c\,\hat{p} (0)}{\sqrt{m^2\,c^2 \,+\, \hat{p}^2 (0)}}\,t\;.
\end{equation}
As also noted in Ref. \cite{Lamme}, the operator $\d \hat{x}/\d t$ is the velocity operator, whose definition has been introduced in quite a 
natural way within the present context, without the necessity of defining a new position operator as usually done in the case of the Dirac 
theory and as discussed in the forthcoming section.

\begin{figure}[htb]
\centering
\includegraphics[height=7cm]{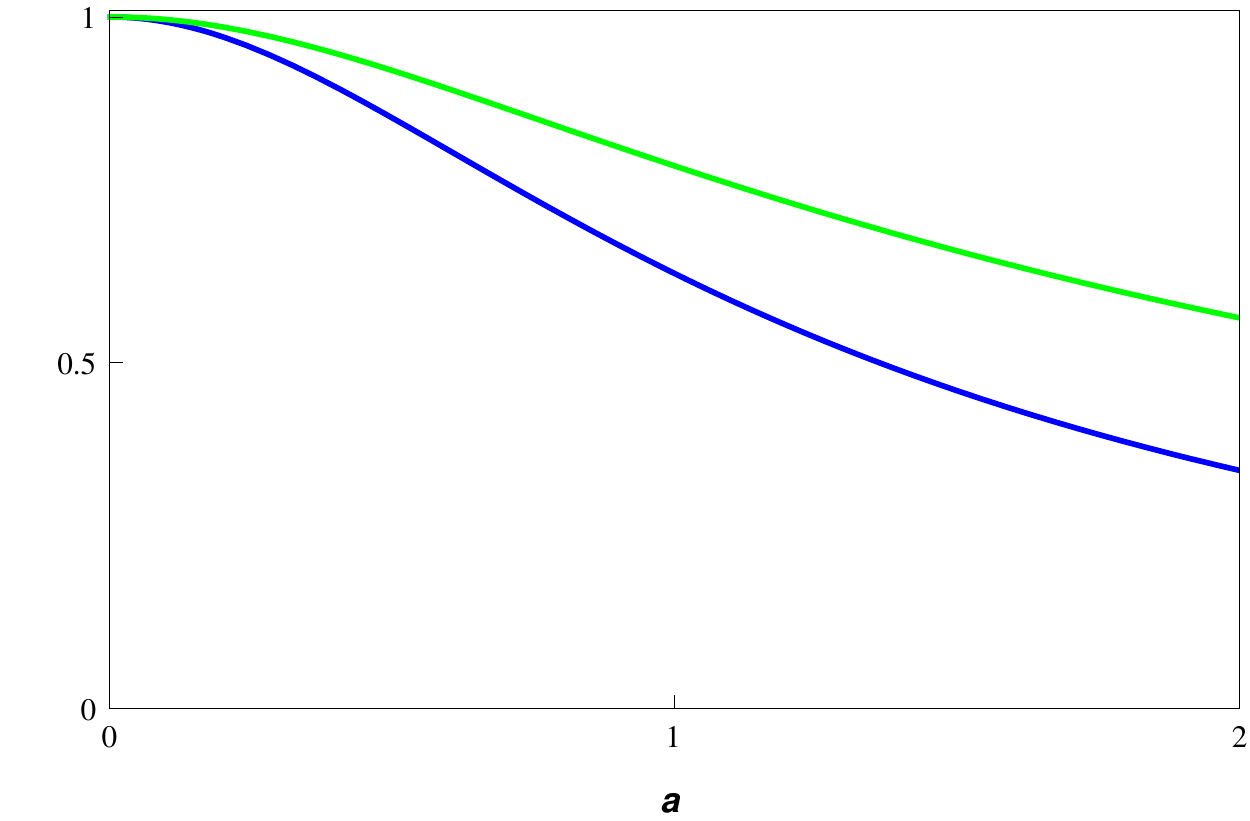}
\caption{Plot of the functions $R (a)$ (blue) and $F (a)$ (green) defined in the text.}
\label{fig:spread} 
\end{figure}

The dynamical behavior of a wave packet undergoing an evolution ruled by the Hamiltonian \eqref{eq:relham} can be inferred from the 
evaluation of the particle position and momentum at a given time $t$. By assuming a packet initially Gaussian 
\begin{equation}
\Psi (x) \,=\, \frac{1}{(2\,\pi\,\sigma^2)^{1/4}}\,\exp\left(-\,\frac{x^2}{4\,\sigma^2}\right)\,,
\end{equation}   
we find that it spreads in time according to the relation
\begin{eqnarray}
\sigma^2 (t) \!\!&=&\!\! \left\langle \hat{x}^2 (0) \,+\, \frac{c^2\,\hat{p}^2 (0)}{m^2\,c^2 \,+\, \hat{p}^2 (0)}\,t^2 \right\rangle \\
                       &=&\!\! \frac{1}{\sqrt{2\,\pi}\,\sigma}\,\int_{-\infty}^\infty\,\d x\,\exp\left(-\,\frac{x^2}{4\,\sigma^2}\right)\,
                                   \left[\hat{x}^2 (0) \,+\, \frac{c^2\,\hat{p}^2 (0)}{m^2\,c^2 \,+\, \hat{p}^2 (0)}\,t^2\right]\,
                                   \exp\left(-\,\frac{x^2}{4\,\sigma^2}\right)\,. \nonumber
\end{eqnarray}
The evaluation of the contribution depending on the momentum operator can be performed using identities \eqref{eq:traide} and 
\eqref{eq:glais}, and yields
\begin{eqnarray}
A(t) \!\!&=&\!\! \left\langle \frac{c^2\,\hat{p}^2 (0)}{m^2\,c^2 \,+\, \hat{p}^2 (0)}\,t^2 \right\rangle \\
           &=&\!\!\frac{c^2\,\lbar_c^2\,t^2}{\sqrt{2\,\pi}\,\sigma}\,\int_{-\infty}^\infty\,\d x\,\exp\left(-\,\frac{x^2}{4\,\sigma^2}\right)\,
                   \partial_x^2\,\left\{\int_0^\infty\,\d s\frac{\e^{-\,s}}{\sqrt{\Sigma (s)}}\,\exp\left(-\,\frac{x^2}{4\,\sigma^2\,\Sigma (s)}\right)\right\}
                   \,, \nonumber
\end{eqnarray}
where
\begin{equation}
\Sigma (s) \,=\, 1 \,+\, a^2\,s\, \qquad\qquad \left(a \,=\, \frac{\lbar_c}{\sigma}\right)\,. \nonumber
\end{equation}
The explicit evaluation of the previous integral allows to write the width of the packet in the form   
\begin{equation}
\label{eq:relspre}
\sigma^2 (t) \,=\, \sigma^2 \left[1 \,+\, \frac{1}{4}\,\left(\frac{a}{\sigma}\right)^2\,R(a)\,c^2\,t^2\right]\,,
\end{equation}
where 
\begin{equation}
\label{eq:rofa}
R (a) 
\,=\,2\,\sqrt{2}\,\int_0^\infty\,\d s\,\frac{\e^{-\,s}}{(2 \,+\, a^2\,s)^{3/2}}
\end{equation}
Equation \eqref{eq:relspre} reduces to its nonrelativistic version for $R (a) = 1$.  As shown in Fig. \ref{fig:spread}, the function $R (a)$ decreases 
with increasing $a$. This means that when the particle is initially localized better than its Compton wavelength, the spreading of the packet 
occurs at a rate slower than in the non relativistic case. Moreover, it is easy to show that 
\begin{equation}
\label{eq:xtx0}
\langle \left[\hat{x} (t),\hat{x} (0)\right] \rangle \,=\, -\,i\,\lbar_c\,F(a)\,c\,t\,,
\end{equation}
with
\begin{equation}
\label{eq:fcomm}
F (a) \,=\, \frac{2\,\sqrt{2}}{\sqrt{\pi}}\,\int_0^\infty\, \d s\,\frac{\sqrt{s}\,\e^{-\,s}}{\sqrt{2 \,+\, a^2\,s}}\;,
\end{equation}
in which the standard nonrelativistic result is corrected by the function $F (a)$. Also this function decreases with increasing argument 
(see Fig. (\ref{fig:spread})), and the conclusion drawn for the spreading extends to the Heisenberg principle.
 
The role of the relativistic corrections in eqs. (\ref{eq:rofa}) and (\ref{eq:fcomm}) can be understood as follows. The indetermination in position 
induces an indetermination in momentum and thus in velocity, which is in turn responsible for the relativistic corrections that we are dealing with. 
The contributions derived from these corrections are, however, of higher order in $\hbar$, as easily inferred from the series expansion of the 
functions $R (a)$ and $F (a)$. For example, the function $R (a)$ at lowest order in $a$ reads 
\begin{equation}
R (a) = 1 \,-\, \frac34\,a^2\;.
\end{equation}
This means that the ordinary nonrelativistic limit is obtained by just keeping the previous results at the lowest order in $a^2$.

In the case of a particle subject to a linear potential 
\begin{equation}
\hat{H} \,=\, c\,\sqrt{m^2\,c^2\,+\,\hat{p}^2}\,-\,\hat{f}\,x\,,
\end{equation}
the equations of motion are written as
\begin{equation}
\frac{\d}{\d t}\,\hat{x} \,=\, c\,\frac{\hat{p}}{\sqrt{m^2\,c^2 \,+\, \hat{p}^2}}\,
\qquad\qquad \frac{\d}{\d t}\,\hat{p} \,=\, \frac{1}{i\,\hbar}\,\left[\hat{p},\hat{H}\right] \,=\, \hat{f}\,,
\end{equation}
and, thus, for the position operator we obtain
\begin{equation}
\hat{x} (t) \,=\, \hat{x} (0) \,+\, \frac{c}{\hat{f}}\,\left[\sqrt{m^2\,c^2 \,+\, (t\,\hat{f} \,+\, \hat{p} (0))^2} \,-\,
                       \sqrt{m^2\,c^2 \,+\, \hat{p}^2 (0)}\right]\,.
\end{equation}
The evaluation of average values becomes more complicated, but still requires the use of fractional operators, we will not further dwell 
on this point but note that the method of the fractional derivatives offers a useful tool to treat problems involving even more complicated 
potentials. Also in this case we have assumed that the Schr\"odinger and Heisenberg pictures are linked by the usual unitary 
transformation (we have indeed evaluated the average value of a given operator as 
$\langle \hat{X} (t) \rangle = \langle \Psi | \exp(i\,\hat{H} t/\hbar)\,\hat{X} | \Psi\rangle$). Even though such an assumption sounds 
trivial, it is not taken for granted.  In particular, when a potential is included, its mathematical legitimacy should be studied in a more 
rigorous way.

\section{Parametrizazion of a unit matrix and fractional operators} \label{sec:param}
In the previous section we have ignored the existence of another procedure to eliminate the square roots, which is essentially the Dirac 
parameterization method in terms of Clifford numbers \cite{Riesz}. The price to be paid is the necessity of introducing a matrix realization 
of these numbers, which inevitably increases the dimensionality of the problem.

Before proceeding further, we it believe worthwhile to treat the Dirac method using a slightly unconventional procedure, according to which we 
will point out that this method is a tool to get the square root of a $2\times2$ or a $4\times4$ unit matrix in terms of Pauli or Clifford numbers, 
respectively. 

The use of the Pauli numbers (matrices)\footnote{From now on, an underlined letter denotes a matrix and a product of 
matrix by a (pseudo)differential operator will be denoted by a hat over an underlined letter.}
\begin{equation}
\underline{\sigma}_1 \,=\, \left(
   \begin{array}{cc}
      0 & 1   \\
      1 & 0  
    \end{array}
\right)\;, \qquad 
\underline{\sigma}_2 \,=\, \left(
   \begin{array}{cc}
      0 & -\,i   \\
      i & 0  
    \end{array}
\right)\,, \qquad 
\underline{\sigma}_3 \,=\, \left(
   \begin{array}{cc}
      1 & 0   \\
      0 & -\,1  
    \end{array}
\right)\,,
\end{equation}
with
\begin{equation}
\left\{\underline{\sigma}_i,\underline{\sigma}_j\right\} \,=\, 2\,\delta_{ij}\,\underline{1}\, \qquad \qquad
\left[\underline{\sigma}_i,\underline{\sigma}_j\right] \,=\, 2\,i\,\epsilon_{ijk}\,\underline{\sigma}_k \qquad\qquad (i,j,k = 1,2,3)\,,
\end{equation}
allows us to estabilish the following identity ($\vec{v} = (v_1,v_2,v_3)$, $v_i \in \mathbb{C}$):
\begin{equation}
\label{eq:pauli}
\underline{N} \,=\, \vec{\underline{\sigma}} \cdot \vec{v} \,=\, v\,\sqrt{\underline{1}}\,.
\end{equation}

The anti-commutative properties of the Pauli numbers allows us also to prove that
\begin{equation}
\underline{N}^{2\,k} \,=\, N^{2\,k}\, \underline{1}\,, \qquad\qquad \underline{N}^{2\,k \,+\, 1} \,=\, N^{2\,k}\, \underline{N}\,,
\end{equation}
from which it is easy to show that
\begin{equation}
\label{eq:expyN}
\e^{y\,\underline{N}} \,=\, \cosh (y\,N)\,\underline{1} \,+\, \frac{1}{N}\,\sinh (y\,N)\,\underline{N}\,.
\end{equation}

As an example, let us apply the parameterization \eqref{eq:pauli} to the case of the relativistic one-dimensional 
Schr\"odinger equation. In this case, the relevant Hamiltonian (see Eq. \eqref{eq:schro} written in terms of the usual coordinate 
$x,t$) can be written as 
\begin{equation}
\label{eq:hamsch}
\hat{\underline{H}} \,=\, \underline{\sigma}_1\,\hat{p}\,c \,+\, \underline{\sigma}_3\,m\,c^2
\end{equation}
and the associated Schr\"odinger equation reads
\begin{equation}
\label{eq:dir2d}
i\,\hbar\,\partial_t\,\underline{\Psi} \,=\,  \left(
   \begin{array}{cc}
      m\,c^2     & \hat{p}\,c \\
      \hat{p}\,c & -\, m\,c^2
    \end{array}
\right)\,\underline{\Psi}\,,
\end{equation}
where $\underline{\Psi}$ is a two-component wave function. This equation should not be confused with the Pauli equation, but it 
should be understood as a two-dimensional Dirac equation with the components of the wave function referring to the positive and 
negative energy states.

By using Eq. \eqref{eq:expyN}, the solution of Eq. \eqref{eq:dir2d} can be written as 
\begin{eqnarray}
\underline{\Psi} (t) \!\!&=&\!\! \exp\left(-\,\frac{i}{\hbar}\,\hat{\underline{H}}\,t\right)\,\underline{\Psi} (0) \nonumber \\
                                 &=&\!\! \left\{\cos \left(\frac{c}{\lbar_c}\,\sqrt{1 \,+\, \hat{\pi}^2}\,t\right)\,\underline{1} \,-\,
                                             i\,\frac{\sin \left(\frac{c}{\lbar_c}\,\sqrt{1 \,+\, \hat{\pi}^2}\,t\right)}{\sqrt{1 \,+\, \hat{\pi}^2}}\,
                                             (\underline{\sigma}_1\,\hat{\pi} \,+\, \underline{\sigma}_3)\right\}\,\underline{\Psi} (0)\, \qquad 
                                             \left(\hat{\pi} \,=\, \frac{\hat{p}}{m\,c}\right)
\end{eqnarray}
that represents a kind of oscillation between the two states, with negative and positive energies. This aspect of the problem can be 
better understood from the Heisenberg equations of motion associated with the Hamiltonian \eqref{eq:hamsch}, namely 
\begin{equation}
\frac{\d}{\d\,t}\,\hat{x} \,=\, c\,\underline{\sigma}_1\, \qquad \frac{\d}{\d\,t}\,\hat{p} \,=\, 0\, \qquad
i\,\hbar\,\frac{\d}{\d\,t}\,\underline{\sigma}_1 \,=\, -\,2\,i\,m\,c^2\,\underline{\sigma}_2\,.
\end{equation}
This equation should be completed with the Hamilton equation of motion for the $\underline{\sigma}_k$ matrices that can be written 
in a vector form as follows:
\begin{equation}
\label{eq:sigma}
\frac{\d}{\d\,t}\,\vec{\underline{\sigma}} \,=\, \vec{\hat{\Omega}}\,\times\,\vec{\underline{\sigma}} \qquad\qquad 
\vec{\hat{\Omega}} \,=\, 2\,\frac{c}{\lbar_c}\,(\hat{\pi}, 0, 1)\,.
\end{equation}
This equation is reminiscent of the Bloch-type equation obtained by Feynman, Vernon, and Hellworth in their rehandling of the 
two-state Schr\"odinger equation \cite{FeyHV}. In this case the two states are represented by the positive and negative energy 
solutions and the coupling between positive and negative states is realized by the matrix $\underline{\sigma}_3$ , whose time 
dependence is just an oscillating function between positive and negative values. In a forthcoming paper we will discuss more 
thoroughly the physical properties of eqs. \eqref{eq:dir2d} and \eqref{eq:sigma}. 

Analogous considerations can be developed for the pseudoheat equation \eqref{eq:psehea}, that, in a two-component form, reads 
\begin{equation}
\partial_ t\,\underline{\Psi} \,=\, -\,\left(
   \begin{array}{cc}
       1   & i\,\partial_x   \\
      i\,\partial_x & -1
    \end{array}
\right)\,\underline{\Psi}\,,
\end{equation}
where the negative and positive solutions are now interpreted as backward and forward heat fluxes. 

In the case of $4 \times 4$ matrices, the Dirac parametrization applies:
\begin{equation}
\label{eq:param}
\underline{N} \,=\, \vec{\underline{\alpha}} \cdot \vec{v} \,+\, \underline{\beta}\,q \,=\, N\,\sqrt{\underline{1}} \qquad\qquad 
(N = \sqrt{v^2\,+\,q^2} \in \mathbb{C}),
\end{equation}
where $\vec{\underline{\alpha}}, \underline{\beta}$ are the Clifford numbers (matrices) realized as
\begin{equation}
\label{eq:cliff}
\underline{\alpha}_1 \,=\, \left(
   \begin{array}{cccc}
      0 & 0 & 0 & 1   \\
      0 & 0 & 1 & 0   \\
      0 & 1 & 0 & 0   \\
      1 & 0 & 0 & 0   
    \end{array}
\right)\,, \quad 
\underline{\alpha}_2 \,=\, \left(
   \begin{array}{cccc}
      0 & 0 & 0 & -\,i   \\
      0 & 0 & i & 0      \\
      0 & -\,i &  0 & 0  \\
      i & 0 & 0 & 0      
    \end{array}
\right)\,, \quad 
\underline{\alpha}_3 \,=\, \left(
   \begin{array}{cccc}
      0 & 0 & 1 & 0    \\
      0 & 0 & 0 & -\,1 \\
      1 & 0 & 0 & 0    \\
      0 & -\,1 & 0 & 0 
    \end{array}
\right)\,, \quad
\underline{\beta} \,=\, \left(
   \begin{array}{cccc}
      1 & 0 & 0 & 0    \\
      0 & 1 & 0 & 0 \\
      1 & 0 & 0 & 0    \\
      0 & 0 & 0 & -\,1 
    \end{array}
\right)\,,
\end{equation}
with
\begin{equation}
\label{eq:anticm}
\left\{\underline{\alpha}_j,\underline{\alpha}_k\right\} \,=\, 2\,\delta_{jk}\,\underline{1}\,, \qquad 
\left\{\underline{\alpha}_j,\underline{\beta}\right\} \,=\, 0\,, \qquad 
\underline{\beta}^2 \,=\, \underline{1} \qquad \quad (j,k = 1,2,3)\;.
\end{equation} 

We stress that, if the only request is to write $\underline{N}$ as the product of a number by the square root of a unit matrix, 
other parametrizations, like
\begin{equation}
\underline{N} \,=\, \vec{\underline{\alpha}}^\prime \cdot \vec{v} \,+\, \underline{\alpha}_3\,u\,, \qquad\qquad 
\vec{\underline{\alpha}}^\prime \,=\,(\underline{\alpha}_1, \underline{\alpha}_2, \underline{\beta})\,,
\end{equation}
or
\begin{equation}
\label{eq:pardia}
\underline{N} \,=\, N\,\underline{\beta}\,, 
\end{equation}
can also be used. The last parametrization has the advantage of being expressed in terms of a diagonal matrix. In some sense  
this is the essence of the so-called Foldy-Woythusen transformation \cite{BjoDr}, and will be discussed in the forthcoming 
section. 

In order to see whether this discussion has any relevance to the theory of fractional derivatives and/or pseudo-operators,  
we consider the following $4 \times 4$ operator:
\begin{equation}
\underline{\hat{O}} \,=\, (1 \,-\, \partial_x^2)\,\underline{1}\,.
\end{equation}
The square root of this operator can be written in terms of the Clifford matrices introduced before as follows:
\begin{equation}
\hat{\underline{O}}^{1/2} \,=\, i\,\vec{\underline{\alpha}} \cdot \vec{\hat{d}} \,+\, \underline{\beta}\, \qquad\qquad
\hat{d}_k \,=\, \frac{1}{\sqrt{3}}\,\partial_x \qquad (k = 1,2,3)\,,
\end{equation}
and therefore the pseudoheat equation \eqref{eq:psehea} can be rewritten in the following form
\begin{equation}
\label{eq:psedir}
\partial_\tau\,\underline{\Phi} \,=\, -\,\hat{\underline{O}}^{1/2}\,\underline{\Phi}
\end{equation}
where $\underline{\Phi}$ is a four-component function. By introducing the matrices
\begin{equation}
\underline{\gamma}_k \,=\, \underline{\beta}\,\underline{\alpha}_k \qquad\qquad (k = 1,2,3),
\end{equation}
and multiplying both sides of Eq. \eqref{eq:psedir} by $\underline{\beta}$, one has ($\tau^\prime = - \tau$)
\begin{equation}
\label{eq:headir}
\underline{\beta}\,\partial_{\tau^\prime}\,\underline{\Phi} \,=\, 
\left(i\,\vec{\underline{\gamma}} \cdot \vec{\hat{d}} \,+\, \underline{1}\right)\,\underline{\Phi}\,,
\end{equation}
that, as it will be discussed in the forthcoming section, is just a Dirac-like form. The last remark suggest that a link can be 
found between fractional calculus and Clifford algebras; further comments about this aspect will be presented in the 
next sections.

\section{Dirac equation} \label{sec:dirac}
In this section we will treat the Dirac equation and its implications for the fractional and pseudodifferential operators calculus. 
By assuming that the parametrization \eqref{eq:param} can be applied also in then case in which $\vec{v}$ is replaced by a 
differential operator, the Dirac equation for a free particle with mass $m$ can be written in the Schr\"odinger form with a  
Hamiltonian given by
\begin{equation}
\underline{\hat{H}} \,=\, c\,\vec{\underline{\alpha}} \cdot \vec{\hat{p}} \,+\, \underline{\beta}\,m\,c^2\,,
\end{equation}
i.e., in terms of the normalized variables introduced in Eq. \eqref{eq:schro}
\begin{equation}
\label{eq:dirac}
i\,\partial_\tau \,\underline{\Psi} \,=\, \left(\vec{\underline{\alpha}} \cdot \vec{\hat{\pi}} \,+\,\underline{\beta}\right)\,\underline{\Psi}\;.
\end{equation}
Moreover, by multiplying both sides of this equation for the matrix $\underline{\beta}$ it assumes the form
\begin{equation}
i\,\underline{\beta}\,\partial_\tau \,\underline{\Psi} \,=\, \hat{\underline{S}}\,\underline{\Psi}
\end{equation}
with
\begin{equation}
\hat{\underline{S}} \,=\, \vec{\underline{\gamma}} \cdot \vec{\hat{\pi}} \,+\,\underline{1} \,=\, \sqrt{1 \,-\, \hat{\pi}^2}\,\sqrt{\underline{1}}\,,
\end{equation}
where the analogy with the pseudo-heat equation \eqref{eq:headir} becomes even closer. The evolution operator associated with 
Eq. \eqref{eq:dirac} is given by 
\begin{equation}
\hat{\underline{U}} (\tau) \,=\, \cos (\sqrt{1 \,+\, \hat{\pi}^2}\,\tau)\,\underline{1} \,-\,
                                             i\,\frac{\sin (\sqrt{1 \,+\, \hat{\pi}^2}\,\tau)}{\sqrt{1 \,+\, \hat{\pi}^2}}\,
                                             (\vec{\underline{\alpha}} \cdot \vec{\hat{\pi}} \,+\, \underline{\beta}^{- 1})\,,
\end{equation}
and the corresponding solution
\begin{equation}
\underline{\Psi} (\eta, \tau) \,=\, \hat{\underline{U}} (\tau)\, \underline{\phi} (\eta)
\end{equation}
shows all the relevant features of the evolution of a free Dirac particle, including the \emph{Zitterbewegung}, if 
$\underline{\phi}$ contains positive and negative energy components.

The Heisenberg equation for the position and momentum operators reads
\begin{equation}
\frac{\d}{\d \tau} \,\vec{\hat{\eta}} \,=\, -\,i\,\left[\vec{\hat{\eta}},\,\vec{\underline{\alpha}} \cdot \vec{\hat{\pi}} 
\,+\, \underline{\beta}\right] \,=\, \vec{\underline{\alpha}}\,, \qquad\qquad \partial_\tau \,\vec{\hat{\pi}} \,=\, 0\,,
\end{equation}
i.e., in this formalism the velocity operator is associated with the Clifford matrices. As for the matrices $\alpha_k$ 
one has
\begin{equation}
\frac{\d}{\d \tau}\, \vec{\underline{\alpha}} \,=\,2\,i\,\left(\vec{\hat{\pi}} \,-\, 
\vec{\underline{\alpha}}\,\hat{\underline{H}}\right)\,,
\end{equation}
that allows to get the solutions for the position operator in the form
\begin{equation}
\vec{\hat{\eta}} (\tau) \,=\, \vec{\hat{\eta}} (0) \,+\, \tau\,\vec{\hat{\pi}}\,\underline{H}^{- 1} \,+\,
\frac{i}{2}\,\underline{H}^{- 1}\,\left(\vec{\underline{\alpha}} (0) \,-\, \vec{\hat{\pi}}\,\underline{H}^{- 1}\right)\,
\left(\e^{-\,2\,i\tau\,\hat{\underline{H}}} \,-\,1\right)\,.
\end{equation}
The first two terms in this equation coincides with Eq. \eqref{eq:xoper}; the last is a further term accounting for the interference 
between particles with positive and negative energy (the \emph{Zitterbewegung} term) and can be removed by their decoupling. 
Being the matrix $\underline{\beta}$ diagonal, the parametrization \eqref{eq:pardia} automatically satisfies such a request, and 
yields 
\begin{equation}
\label{eq:soleta}
\frac{\d}{\d \tau}\, \vec{\hat{\eta}} \,=\, \underline{\beta}\,\frac{\vec{\hat{\pi}}}{\sqrt{1 \,+\, \hat{\pi}^2}}\,.
\end{equation}
Since the eigenvalues of $\underline{\beta}$ have the values $\pm\,1$, the positive and negative energy solutions have velocities 
in opposite directions. By integrating Eq. \eqref{eq:soleta}, since the momentum remains constant, we get 
\begin{equation}
\vec{\hat{\eta}} (\tau) \,=\, \vec{\hat{\eta}} (0) \,+\, \underline{\beta}\,\frac{\vec{\hat{\pi}}}{\sqrt{1 \,+\, \hat{\pi}^2}}\,\tau\,.
\end{equation}

We close this section by addressing the problem of the existence of alternative forms for the Dirac equation depending on the 
chosen parametrization. In 1971, in a not widespread known paper \cite{Dira2}, Dirac himself proposed a four-component equation, 
not containing negative energy solutions and valid for particles with integer spin. This equation, written in a form closely similar to 
the electron relativistic case, contains a different realization of the matrices $\underline{\alpha}_k$ and $\underline{\beta}$. Without 
entering into the details of the physical implications of this type of equation, which will be discussed elsewhere, we consider the 
problem from the point of view that inspired this paper, by noting that for the product of the square root of a number the unit matrix 
also the following parametrization applies
\begin{equation}
\underline{N} \,=\, \vec{\underline{\kappa}} \cdot \vec{w} \,+\, i\,\underline{\delta}\,r \qquad\qquad 
(N = \sqrt{w^2\,+\,r^2} \in \mathbb{C})\,.
\end{equation}
where $\underline{\kappa}_1 = - \underline{\alpha}_3$,  $\underline{\kappa}_2 = \underline{\alpha}_1$, 
$\underline{\kappa}_3 = \underline{\beta}$, and
\begin{equation}
\underline{\delta} \,=\, \left(
   \begin{array}{cccc}
      0 & 0 & 1 & 0   \\
      0 & 0 &  0 & 1   \\
     -1 & 0 & 0 & 0   \\
      0 & -1 & 0 & 0   
    \end{array}
\right)\,.
\end{equation}
These matrices satisfy the same anticommutation relation of Eq. \eqref{eq:anticm}, except that $\underline{\delta}^2 \,=\, -\,\underline{1}$. 
Accordingly we can introduce the following equation 
\begin{equation}
i\,\partial_\tau \,\underline{\Psi} \,=\, (\vec{\underline{\kappa}} \cdot \vec{\hat{\pi}} \,+\, i\,\underline{\delta})\,\underline{\Psi}\,.
\end{equation}                                                             
It resembles the ordinary Dirac equation, but the associated Hamiltonian is a non-Hermitian operator. Although it has been derived using 
the correct correspondence with the relativistic Schr\"odinger Hamiltonian, it cannot be exploited in unitary evolution processes and its 
physical properties are completely different from the ordinary Dirac equation (see \cite{Datt1} for further comments on the use of the 
matrices $\underline{\kappa}_j$ and $\underline{\delta}$ in the derivation of Dirac-like matrices).

It is also interesting to note that
\begin{equation}
\underline{N} \,=\, \vec{\underline{\kappa}} \cdot \vec{w} \,+\, \underline{\delta}\,r 
\end{equation}
is such that
\begin{equation}
N = \sqrt{w^2\,-\,r^2}
\end{equation}
and therefore it can provide the square root of the matrix $- \underline{1}$ or even of the null matrix (when $w^2 = r^2$).

\section{Conclusions}\label{sec:conclus}
In this paper we have discussed a number of problems showing how different analytical methods, from apparently uncorrelated fields, 
can be merged to provide interesting and useful tools. We have presented a general view on the theory of pseudo-operators and of 
fractional derivatives and on their implications for problems regarding the relativistic heat and the Dirac equations. The latter case has 
been discussed adopting a method involving the extraction of the square root of the unit matrix. We have touched many points and each 
of them would deserve a separate and deeper treatment. In particular, we have stressed that alternative parametrization may lead to a 
meaningful alternative. This observation is not entirely new and, apart from the already quoted Dirac paper \cite{Dira2}, the same problem 
was addressed also in Ref. \cite{BieHvD}. In this paper the authors insist on what they call the Dirac ``dichotomy'' and show that the 
alternative derivation, based on a $2 \times 2$ matrix representation, leads to a two-component equation of the type discussed in 
Sec. \ref{sec:param}.

In the paper we have not discussed the problem of the discrete symmetries of Eq. \eqref{eq:dir2d}. For differential equations with 
fractional derivatives this problem deserves particular care  \cite{Vasqu1}. We only remark that for parity ($P$) and time reversal ($T$), the 
corresponding operators can be constructed in terms of the Pauli matrices in analogy to the procedure followed in Ref. \cite{Nierd}. 

A further point, which has not been touched in the paper,  is the nature of equations like the relativistic heat equation \eqref{eq:psehea}, 
generically treated as an evolution equation. The square root operator does not allow a classification in the usual sense (elliptic, parabolic, hyperbolic) 
(see Ref. \cite{Vasqu2}). We have avoided this point, since we have dealt with a genuine Cauchy problem, namely an initial value problem 
concerning the solution of a differential equation first order in time. The techniques associated with the evolution operator method are therefore 
sufficient, provided that the operator is well defined. The inclusion of boundary conditions would imply mathematical problems and technicalities 
beyond the scope of the paper.

We close the paper briefly mentioning the application of this method to the case of the wave propagation. It is well known that the 
non-relativistic Schr\"odinger equation is used in classical optics to treat the paraxial wave propagation, namely the case in which the 
propagation occurs without significant variation of the field with respect to the axis of propagation \cite{LeoFoc}. The equation is obtained as a 
consequence of the application of the so-called Leontovich-Fock approximation to the Helmholtz equation \cite{Komar}. The Hamiltonian 
describing the propagation of an optical ray in a medium with an index of refraction $n$ is\footnote{This Hamiltonian is not invariant under 
Lorentz transformations and therefore cannot be considered relativistic. Nevertheless, a Dirac-like parametrization can be performed as
$$
\hat{\underline{H}} \,=\, -\,\vec{\underline{\alpha}} \cdot \vec{\hat{p}} \,+\, i\,\underline{\beta}\,n\,.
$$
} 
\begin{equation}
H \,=\, -\,\sqrt{n^2 \,-\, p^2}\,.
\end{equation} 
If the medium is homogeneous $n$ can be considered constant; the variable $p$ denotes the beam divergence and the use of the standard 
procedure allows us to replace it with the operator $-\,i\,\lbar\,\nabla_q$, where $q$ is the transverse coordinate and  $\lambda$ is the 
wavelength of the propagating radiation. The ``quantization'' procedure, involving $\lbar$ instead of $\hbar$, is guaranteed by the fact that 
$p$ and $q$ are canonically conjugated variables. The method we have described in this paper can therefore be applied to the study of 
nonparaxial wave evolution and to the quantitative analysis of the deviation from the paraxial case.
\appendix

\section{}
The identity 
\begin{equation}
\label{eq:doetra}
\e^{-\,x\,\sqrt{y}} \,=\, \frac{1}{2\,\sqrt{\pi}}\,\int_0^\infty\,\d t\, t^{-\,3/2}\,\exp\left(-\,\frac{1}{4\,t} \,-\, t\,x^2\,y\right)
\end{equation}
is sometimes referred to as the Doetsch integral transform \cite{Doets}. We dwell on the proof of this identity because, albeit available 
in literature \cite{Andre, BaDat}, it yields a first glimpse of how methods from different fields of calculus can be combined to get the 
desired result.   

The substitution $t = 1/\xi^2$ in the integral in the right-hand side of Eq. \eqref{eq:doetra} leads us to consider the function defined as
\begin{equation}
\label{eq:intIab}
I(a,b) \,=\, \frac{1}{\sqrt{\pi}}\,\int_0^\infty\,\d\xi \,\exp\left(-\,a^2\,\xi^2 \,-\, \frac{b^2}{\xi^2}\right)\,,
\end{equation}
which is easily shown to satisfy the differential equation
\begin{equation}
\partial_b\,I(a,b) \,+\,2\,\,a\,I(a,b) \,=\, 0\,, \qquad\qquad I(a,0) \,=\, \frac{1}{2\,a}\,.
\end{equation}
The solution of this equation is given by
\begin{equation}
\label{eq:solIab}
I(a,b) \,=\, I(a,0)\,\e^{-\,2\,a\,b}\;,
\end{equation}
that for $a = 1/2$, $b = x\,\sqrt{y}$ reproduces Eq. \eqref{eq:doetra}. In this case, from Eq. \eqref{eq:solIab} and \eqref{eq:intIab} we 
obtain also the identity
\begin{equation}
\e^{-\,x\,\sqrt{y}} \,=\,  \frac{1}{\sqrt{\pi}}\,\int_0^\infty\,\d\xi \,\exp\left(-\,\frac{\xi^2}{4} \,-\, \frac{x^2\,y}{\xi^2}\right)\,,
\end{equation}
that can be exploited to write the solution of Eq. \eqref{eq:sqrder} in the following alternative form
\begin{eqnarray}
F(x,\tau) \!\!&=&\!\! \frac{1}{\sqrt{\pi}}\,\int_0^\infty\,\d\xi \,\exp\left(-\,\frac{\xi^2}{4} \,-\, \frac{\tau^2}{\xi^2}\,\partial_x\right)\,f(x)
                              \nonumber \\
                  &=&\!\! \frac{1}{\sqrt{\pi}}\,\int_0^\infty\,\d\xi \,\e^{\xi^2/4}\,f\left(x \,-\, \frac{\tau^2}{\xi^2}\right)\,.
\end{eqnarray}

\section{}
In this appendix we show how the already quoted Laplace transform method can be applied to evaluate the square root of a matrix. 

As an example, let us consider the following matrix
\begin{equation}
\underline{R} \,=\, a\,\underline{1} \,+\, b\,\underline{\sigma}_1\;,
\end{equation} 
with $a > b > 0$. For its square root we can write
\begin{equation}
\sqrt{\underline{R}} \,=\, \frac{\underline{R}}{\sqrt{\underline{R}}} \,=\, \underline{R}\,\left(\frac{1}{\sqrt{\pi}}\,\int_0^\infty\,\d s\,
                                       \frac{\e^{-\,s\,\underline{R}}}{\sqrt{s}}\right)\,,
\end{equation}
where identity \eqref{eq:traide} has been used, and, as a consequence of Eq. \eqref{eq:expyN}, it turns out that
\begin{eqnarray}
\sqrt{\underline{R}} \!\!&=&\!\! \frac{\underline{R}}{\sqrt{\pi}}\,\int_0^\infty\,\d s\, \frac{\e^{-\,s\,a\,\underline{1}}}{\sqrt{s}}\, 
                                                \left\{\cosh (b\,s)\,\underline{1} \,-\, \sinh (b\,s)\,\underline{\sigma}_1\right\} \nonumber \\
                                   &=&\!\! \frac{\underline{R}}{2}\,\left\{\frac{1}{\sqrt{a \,-\, b}}\,(\underline{1} \,-\, \underline{\sigma}_1) \,+\, 
                                                \frac{1}{\sqrt{a \,+\, b}}\,(\underline{1} \,+\, \underline{\sigma}_1)\right\}\,.
\end{eqnarray}

This method can be applied to more general matrix forms, by combining the previous integral transform with more conventional tools, like 
the Cayley-Hamilton theorem. Moreover, it is not limited to the square root but can be formulated in a wider context. For $\nu > 0$, we get 
indeed 
\begin{equation}
\underline{R}^{-\,\nu} \,=\, \frac{1}{2}\,\left\{\frac{1}{(a \,-\, b)^\nu}\,(\underline{1} \,-\, \underline{\sigma}_1) \,+\, 
                                          \frac{1}{(a \,+\, b)^\nu}\,(\underline{1} \,+\, \underline{\sigma}_1)\right\}\,.
\end{equation}




\begin{thebibliography}{99}

\bibitem{Zasla}
G. M. Zaslawsky, 
Phys. Rept. {\bf 371}, 461 (2002).

\bibitem{Compte}
A. Compte, 
Phys. Rev. E {\bf 53}, 4191 (1996).

\bibitem{Bonil}
B. Bonilla, M. Rivero, and J. J. Trujillo, 
Appl. Math. Comput.  {\bf187}, 68 (2007). 

\bibitem{Taylo}
M. E. Taylor, 
\textit{Pseudo differential Operators} (Princeton University Press, Princeton, NJ, 1981); 
M. A. Shubin, 
\textit{Pseudo differential Operators and Spectral Theory} (Springer-Verlag, Berlin, 1987); 
F. Treves, 
\textit{Introduction to Pseudo differential and Fourier Integral Operators} (Plenum, New York, 1981).

\bibitem{DitPru}
V. A. Ditkin and A. P. Prudnikov, 
\textit{Integral Transforms and Operational Calculus} (Pergamon, New York, 1965).

\bibitem{Dira1}
P. A. M. Dirac, 
Proc. R. Soc. London, Ser. A {\bf 117}, 610 (1928);  {\bf 118}, 351 (1928).

\bibitem{Doets}
G. Doetsch, 
\textit{Handbuch der Laplace Transformation} (Birkhauser, Basel, 1950-1956).

\bibitem{OldSpa}
K. B. Oldham and J. Spanier, 
\textit{The fractional calculus} (Dover, New York, 2002).

\bibitem{Datto}
G. Dattoli, P. L. Ottaviani, A. Torre, and L. V\'azquez, 
Riv. Nuovo Cimento {\bf 20} (2), 1 (1997).

\bibitem{Datt1}
G. Dattoli, S. Khan, and P. E. Ricci, 
Integr. Transf. Spec. F {\bf 19}, 1 (2008).

\bibitem{Datt2} 
G. Dattoli, M. Quattromini, and A. Torre, 
Nuovo Cimento B {\bf 114}, 693 (1999).

\bibitem{BjoDr}
J. D. Bjorken and S.  Drell, 
\textit{Relativistic Quantum Mechanics} (McGraw-Hill, New York, 1964).

\bibitem{Lamme}
C. Lammerzahl, 
J. Math. Phys. {\bf 34}, 3918 (1983).

\bibitem{Datt3}
G. Dattoli, 
J.  Comput.  Appl. Math. {\bf 118}, 111 (2000).

\bibitem{Riesz} 
M. Riesz, 
\textit{Clifford Numbers and Spinors}, Fundamental Theories of Physics Vol. 54, 
(Kluwer, Dordrecht, 1993).

\bibitem{FeyHV}
R. P. Feynman, R. W. Hellwarth, and F. L. Vernon, 
J. Appl. Phys. {\bf 28}, 49 (1957).

\bibitem{Dira2} 
P. A. M. Dirac, 
Proc. R. Soc. London, Ser. A {\bf 322}, 435 (1971).

\bibitem{BieHvD} 
L. C. Biedenharn, M. Y. Han, and H. Van Dam, 
Phys. Rev. D {\bf 6}, 500 (1972).

\bibitem{Vasqu1}
see e. g. L. V\'azquez, \textit{Fractional Equations with Internal Degrees of Freedom}, 
Symposium on Applied Fractional Calculus Industrial Engineering School, University 
of Extremadura, Badajoz, October 15-17, 2007.

\bibitem{Nierd}
J. Nierderle and A. Nikitin, 
J. Nonlinear Math. Phys. {\bf 4}, 436 (1997).

\bibitem{Vasqu2}
L. V\'azquez , R. Vilela Mendes, Appl. Math. Comput. {\bf141}, 125 (2003);
T. Pierantozzi and L. V\'azquez, J. Math. Phys. {\bf 46}, 113512 (2005). 

\bibitem{LeoFoc}
M. A. Leontovich and V. A. Fock, 
Zh. Eksp. Teor. Fiz. {\bf 16}, 557 (1946) [J. Phys USSR {\bf 10}, 13 (1946)].

\bibitem{Komar} 
see e. g. \textit{Classical and Quantum effects in Electrodynamics}, 
Proceedings of the Lebedev Physics Institute Academy of Science of the USSR, 
vol. 176, A. A. Komar ed. (Nova Science, Moscow, 1988).

\bibitem{Andre}
L. C. Andrews, 
\textit{Special functions for Engineers and Applied Mathematicians} (MacMillan, New York, 1985).

\bibitem{BaDat} 
D. Babusci, G. Dattoli, and M. Del Franco, 
\textit{Lectures on Mathematical Methods for Physics}, 
Internal Report No. ENEA RT/2010/5837.



\end{thebibliography}
\end{document}